\documentclass[%
 reprint,
superscriptaddress,
 amsmath,amssymb,
 aps,
floatfix,
]{revtex4-1}

\usepackage{graphicx}
\usepackage{subcaption}
\usepackage{dcolumn}
\usepackage{bm}
\usepackage{braket}
\usepackage{amsmath}
\usepackage{mathtools}
\usepackage[colorlinks,allcolors=blue]{hyperref} 
\usepackage{hyperref}

\begin{document}

\raggedbottom

\preprint{APS/123-QED}

\title{
Low-Temperature 
Annihilation Rate for Quasi-Localized Excitons in Monolayer MoS\textsubscript{2}}

\author{Eric Chatterjee}
\email{echatter@stanford.edu}
\affiliation{Edward L. Ginzton Laboratory, Stanford University, Stanford, California 94305, USA}
\author{Daniel B. S. Soh}
\affiliation{Edward L. Ginzton Laboratory, Stanford University, Stanford, California 94305, USA}
\affiliation{Sandia National Laboratories, Livermore, California 94550, USA}
\author{Christopher Rogers}
\affiliation{Edward L. Ginzton Laboratory, Stanford University, Stanford, California 94305, USA}
\author{Dodd J. Gray}
\affiliation{Edward L. Ginzton Laboratory, Stanford University, Stanford, California 94305, USA}
\author{Hideo Mabuchi}
\email{hmabuchi@stanford.edu}
\affiliation{Edward L. Ginzton Laboratory, Stanford University, Stanford, California 94305, USA}

\date{\today}

\begin{abstract}
The strong Coulomb forces in monolayer transition metal dichalcogenides ensure that optical excitation of band electrons 
gives rise to Wannier-Mott excitonic states, each of which can be conceptualized as a composite of a wavepacket corresponding to center-of-mass motion and an orbital state corresponding to the motion of the electron and hole about the center-of-mass. Here, we show that at low temperature in monolayer MoS\textsubscript{2}, given quasi-localized excitons and consequently a significant inter-exciton spacing, the excitons undergo dipole-dipole interaction and annihilate one another in a manner analogous to Auger recombination. To design our model, we assume that each exciton is localized in a region whose length is on the same scale as the excitonic diameter, thus causing the exciton to behave in a fermion-like manner, while the distance between neighboring excitons is much larger than the exciton diameter. We construct the orbital ladder operators for each exciton and apply Fermi's Golden Rule to derive the overall recombination rate as a function of exciton density.
\end{abstract}

\pacs{Valid PACS appear here}
\maketitle


\section{Introduction}
\par
Recent advances in nanophotonic signal processing have focused increasing attention on novel materials that could provide strong and/or novel nonlinear optical properties in near-planar structures~\cite{Nanophotonics1, Nanophotonics2, Nanophotonics3, Nanophotonics4, Nanophotonics5, Nanophotonics6, Nanophotonics7}. Monolayer transition-metal dichalcogenides (TMDs) are of particular interest in this regard because of their direct band gaps~\cite{MoS2DirectBandGap, MoS2Spectroscopy, MoS2DirectGapDFT, MoS2Luminescence, MoSe2DirectBandGap, WX2DirectBandGap} and substantial optical nonlinearity relative to conventional bulk materials~\cite{MoS2Nonlinearity, MX2Nonlinearity, WS2SaturableAbsorption, MoS2Susceptibilities, OpticalNonlinearitiesMoS2}. The primary nonlinear susceptibilities of monolayer TMDs are excitonic in nature, including Kerr-type nonliearity associated with multiphoton transitions among exciton internal states~\cite{OpticalNonlinearitiesMoS2} and saturable absorption-type nonlinearity associated with exciton-exciton interactions at high density. Here we present a theoretical study connected to the latter effect, which can in principle provide a basis for optical bistability and thus all-optical switching~\cite{Lugiato1983}.
\par
Electron-hole binding energies in monolayer TMDs range from 0.3 eV to 1 eV \cite{ExcitonBindingEnergy, ExcitonBindingEnergy2, ExcitonBindingEnergyWX2, ExcitonBindingEnergyWS2, ExcitonBindingEnergyMoSe2}, resulting in the formation of Wannier-Mott excitons in such materials. One important topic of research involves the calculation of the excitonic decay rates through various channels, which play an essential role in determining the dynamic optical response of a monolayer TMD near the exciton resonance. To this end, recent theoretical and phenomenological studies have derived the radiative loss rate for excitons in MoS\textsubscript{2} \cite{RadiativeDecayMoS2, OpticalNonlinearitiesMoS2} and for other TMDs \cite{ExcitonRadiativeTMDC, ExcitonicLinewidthTMDC}, as well as for other low-dimensional systems such as plasmons in graphene nanoribbons \cite{RadiativeDecayNanoribbons}. In addition, the nonradiative decay rate due to exciton-phonon scattering has been numerically and phenomenologically derived as a function of temperature for WS\textsubscript{2} and MoSe\textsubscript{2} \cite{ExcitonicLinewidthTMDC}.
\par
At sufficiently low temperatures in high quality crystals, the nonradiative decay consisting of exciton-phonon interactions is minimal (due to lack of significant phonon population) compared to the exciton-exciton interaction. This interaction can potentially take the form of either annihilation (via a process analogous to Auger recombination, in which Coulomb interaction between neighboring excitons causes one to be annihilated and the other to be excited) or scattering (direct or exchange). Given two bright ground-state excitons spaced significantly apart, only the recombination process will occur (the reasoning for this will be discussed later). Intuitively, the rate of this decay process should vary with the excitonic spatial density, which can be modulated via the intensity of the incident electromagnetic field. On the theoretical front, Auger recombination of excitons has been previously analyzed in tightly confined 1D systems such as carbon nanotubes \cite{ExcitonAnnihilation1D}.  Experimentally, recent measurements have demonstrated the presence of rapid exciton-exciton annihilation in monolayer MoS\textsubscript{2} \cite{RapidExcitonAnnihilationMoS2}, as well as in other monolayer TMDs such as WSe\textsubscript{2} \cite{RapidExcitonAnnihilationWSe2} and MoSe\textsubscript{2} \cite{RapidExcitonAnnihilationMoSe2}. Here, we will analytically derive the corresponding rate in monolayer MoS\textsubscript{2} as a function of the exciton density, in the quasi-localized regime. Although most recent experiments with monolayer TMDs would seem to take place outside the quasi-localized regime, we note that our results nonetheless appear to be consistent with measured exciton annihilation rates. We discuss interpretations of this apparent agreement in the Discussion.
\par
This paper is organized as follows: In Section~\ref{sec: Exciton-Exciton Interaction Energy}, we derive the exciton-exciton coupling energy as a function of the relative electron to hole positions for a pair of excitons, under the assumption that the spacing between neighboring excitons is much greater than their diameters. This supposition is supported by the fact that for MoS\textsubscript{2}, the ground state excitonic diameter is about $0.67 \textrm{ nm}$ \cite{OpticalNonlinearitiesMoS2}, whereas the experimental data shows rapid Auger recombination even at a nearest-neighbor exciton spacing greater than $10 \textrm{ nm}$ \cite{RapidExcitonAnnihilationMoS2}, while measurements on MoSe\textsubscript{2} (a TMD with properties similar to MoS\textsubscript{2}) demonstrate a saturation of the excitonic density at a nearest-neighbor spacing of $4 \textrm{ nm}$ \cite{RapidExcitonAnnihilationMoSe2}. In Section~\ref{sec: Matrix Elements and Interaction Hamiltonian}, we calculate the matrix elements of the hole-to-electron vectors in the basis of the respective ladders of orbital states and thus obtain the exciton-exciton interaction Hamiltonian in terms of the ladder operators for the two excitons. In Section~\ref{sec: Calculating the Annihilation Rate}, we apply Fermi's Golden Rule to derive the annihilation rate as a function of excitonic density. Finally, in Section~\ref{sec: Results and Discussion}, we evaluate these expressions and discuss the results. Specifically, we focus on the effect of the localization of the excitonic centers-of-mass and the saturation density of excitons on the annihilation rate.

\section{Exciton-Exciton Interaction Energy} \label{sec: Exciton-Exciton Interaction Energy}
\par
We set up this problem by separately considering the coupling of the electron and the hole, respectively, with another exciton. 
\begin{figure}
    \centering
    \includegraphics[width=\columnwidth]{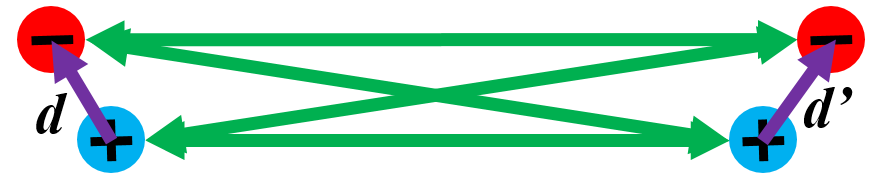}
    \caption{Diagram of the two interacting excitons. The green arrows depict the Coulomb coupling between the two excitons that generates the interaction energy, and the purple arrows depict the hole-to-electron vectors for the excitons ($\bm{d}$ and $\bm{d'}$, respectively). Note that the inter-exciton spacing is much greater than the size of each exciton.}
    \label{fig: two excitons}
\end{figure}
As shown in Fig.~\ref{fig: two excitons}, we apply the assumption that the spacing between neighboring exciton centers is much larger than the hole-electron distance for each exciton. Since the hole-electron distance resembles the monolayer MoS\textsubscript{2} film thickness of $0.65 \textrm{ nm}$ \cite{MoS2Thickness1, MoS2Thickness2}, the Rytova-Keldysh potential energy at the range of the exciton-exciton spacing can be considered to approximately equal the 3D Coulomb potential energy \cite{Rytova, Keldysh}. Labeling the constant factor in the Coulomb energy as $C$, the electron-exciton potential energy is written in terms of spatial coordinates as follows:
\begin{equation}
\begin{gathered}
V_{e-ex} = C\Bigg(\frac{1}{|\bm{r_e}-\bm{r_{e'}}|} - \frac{1}{|\bm{r_e}-\bm{r_{h'}}|}\Bigg).
\end{gathered}
\end{equation}
The hole-exciton potential energy is similar, except with the electron coordinate $\bm{r_e}$ replaced by hole coordinate $\bm{r_h}$ and the sign of the energy flipped since the hole carries a positive charge:
\begin{equation}
\begin{gathered}
V_{h-ex} = -C\Bigg(\frac{1}{|\bm{r_h}-\bm{r_{e'}}|} - \frac{1}{|\bm{r_h}-\bm{r_{h'}}|}\Bigg).
\end{gathered}
\end{equation}
Since the spacing between electron and hole for each exciton is minimal compared to the distance between the excitons, we can use the well-known dipole-dipole approximation to determine the sum of the above two potential energies \cite{DipoleInteraction}. We label the exciton-exciton vector as $\bm{\Delta r} = \bm{r_{CM}} - \bm{r_{CM'}}$ (where $\bm{r_{CM}}$ and $\bm{r_{CM'}}$ represent the center-of-mass positions of the excitons) and the hole-to-electron vector for the unprimed (primed) exciton as $\bm{d} = \bm{r_e} - \bm{r_h}$ ($\bm{d'} = \bm{r_{e'}} - \bm{r_{h'}}$). We find that the interaction energy varies linearly with both $\bm{d}$ and $\bm{d'}$, as desired:
\begin{equation} \label{eq: exciton-exciton potential energy}
\begin{gathered}
V_{ex-ex} = -\frac{C}{(\Delta r)^3} \bigg(\frac{3 \bm{\Delta r} \cdot \bm{d}}{(\Delta r)^2} \bm{\Delta r} - \bm{d} \bigg) \cdot \bm{d'}.
\end{gathered}
\end{equation}
Labeling the ratio of electron mass to total mass as $C_e = \frac{m_e}{m_e + m_h}$ and the corresponding ratio for the hole as $C_h = \frac{m_h}{m_e + m_h}$ (where $m_e$ and $m_h$ are positive constants denoting the electron and hole masses, respectively), we note that $\bm{r_{CM}} = C_e \bm{r_e} + C_h \bm{r_h}$ and $\bm{r_{CM'}} = C_e \bm{r_{e'}} + C_h \bm{r_{h'}}$.
\par
It is worth analyzing $\bm{\Delta r}$, $\bm{d}$, and $\bm{d'}$, since we will promote these variables to operator form later. In general, the Hilbert space spanned by the plane wave states of the two constituent charge carriers (electron and hole) of an exciton is also spanned by a tensor Hilbert space of the center-of-mass degree of freedom and the orbital degree of freedom. This is mathematically demonstrated by the fact that we can express the Schrodinger equation for a single exciton in both of these position bases. The following is the Schrodinger equation in the electron-hole position basis:
\begin{equation} \label{eq: single exciton Hamiltonian first basis}
\begin{gathered}
H_{ex} = -\frac{\hbar^2}{2 m_e} \nabla_e^2 - \frac{\hbar^2}{2m_h} \nabla_h^2 + V_{int}(\bm{r_e} - \bm{r_h}).
\end{gathered}
\end{equation}
Here, $\nabla_e$ and $\nabla_h$ represent the spatial gradients with respect to $\bm{r_e}$ and $\bm{r_h}$, respectively. The convenience of converting the elctron-hole basis to the basis consisting of the center-of-mass position and the hole-to-electron vector derives from the fact that the electron-hole electrostatic potential energy $V_{int}$ is specifically a function of the displacement between the electron and hole positions while being otherwise invariant in the individual positions themselves. In that second basis, the Hamiltonian is therefore separable into a center-of-mass part $H_{CM}$ and an orbital part $H_{orb}$:
\begin{equation}
\begin{split}
H_{ex} &= -\frac{\hbar^2}{2M} \nabla_{CM}^2 - \frac{\hbar^2}{2\mu} \nabla_{orb}^2 + V_{int}(d) \\
&= H_{CM} + H_{orb}.
\end{split}
\end{equation}
Here, $\nabla_{CM}$ and $\nabla_{orb}$ denote the spatial gradients with respect to $\bm{r_{CM}}$ and $\bm{d}$, respectively, and $M$ and $\mu$ represent the total electron-hole mass and reduced electron-hole mass, respectively. Note that $V_{int}$ is entirely included in the orbital part $H_{orb}$, thus ensuring the separability of the Hamiltonian.
\par
Introducing interaction between excitons, we find that the Schrodinger equation for an unprimed exciton and a primed exciton interacting with each other takes the following form:
\begin{equation}
\begin{gathered}
H = H_{ex} + H_{ex'} + H_{ex-ex}.
\end{gathered}
\end{equation}
Here, $H_{ex-ex}$ denotes the perturbation to the total Hamiltonian due to exciton-exciton interaction and is given by the interaction energy from Eq.~\eqref{eq: exciton-exciton potential energy}:
\begin{equation}
\begin{gathered}
H_{ex-ex} = V_{ex-ex}(\bm{r_{CM}}-\bm{r_{CM'}},\bm{d},\bm{d'}).
\end{gathered}
\end{equation}
For the Auger process, the fact that we start with two bright excitons from the lowest excitonic state (since both were optically excited to that level) implies that we finish with a higher-energy bright exciton. As a result, we only consider cases for which the center-of-mass state is centered at zero momentum. If the excitonic center-of-mass were fully delocalized, this would reduce the composite wavefunction to only the orbital part, multiplied by the normalization factor $1/\sqrt{A_{beam}}$, where $A_{beam}$ represents the cross-sectional area of the beam generating the excitons. In reality, defects will constrain the diffusion range \cite{ExcitonDefect,ExcitonDefect2}, and quantum dots can also be artifically synthesized in order to generate the same effect (as will be discussed in Section~\ref{sec: Results and Discussion}). We will solve the annihilation rate specifically for the case in which each exciton is localized to a unique region (the valid range of localization areas will be analyzed in Section~\ref{sec: Results and Discussion}). 
\par
\begin{figure}
    \centering
    \includegraphics[width=\columnwidth]{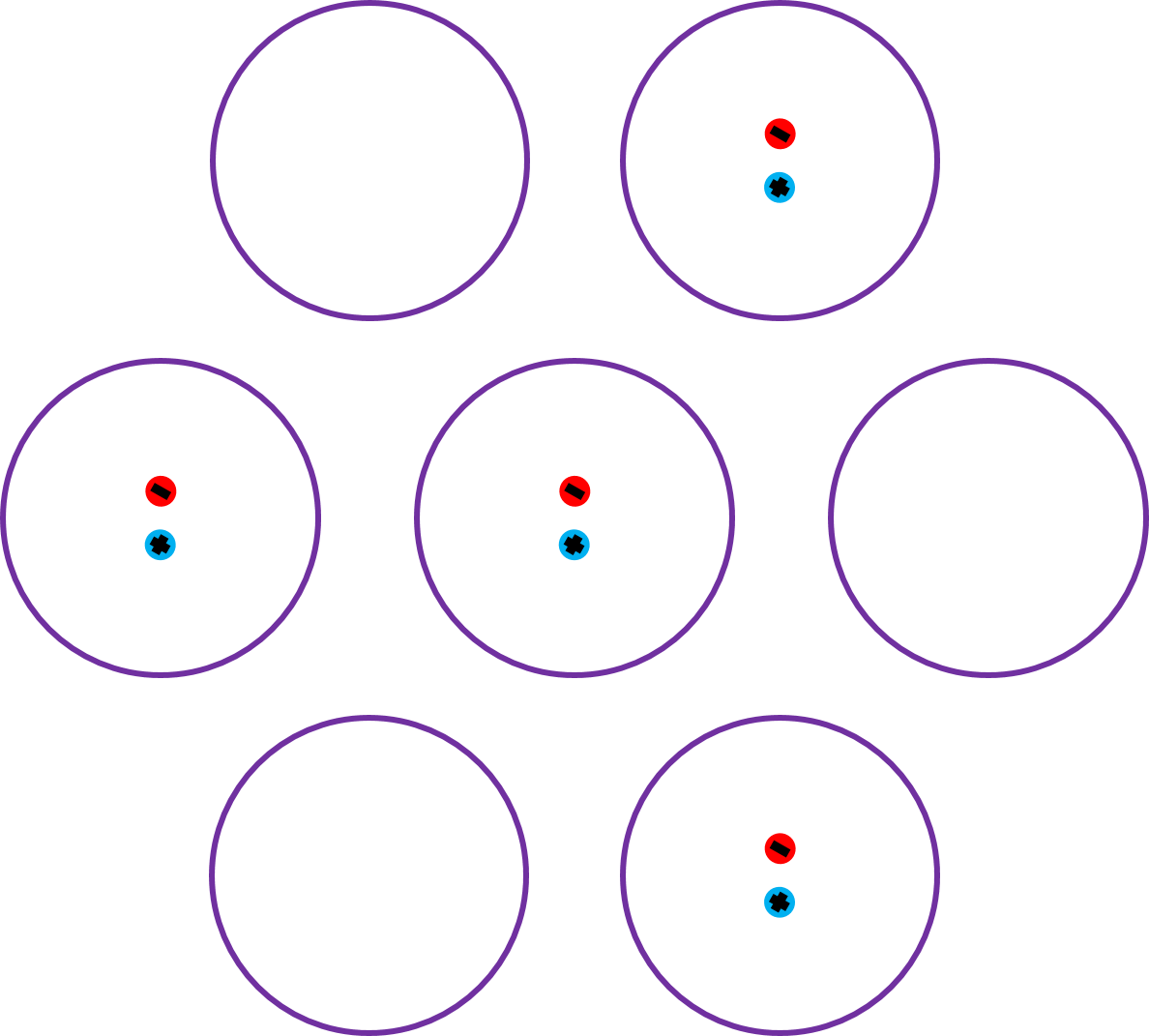}
    \caption{Visualization of the layout of the excitonic regions in position space given partial filling. Each circle forms the boundary of a localization region, which contains either 0 or 1 exciton. The area of each region will be labeled as $A$, the density of regions as $\sigma_r$, and the excitonic density as $\sigma$. Note that if $\sigma < \sigma_r$, then the occupation probability for a given region is $\frac{\sigma}{\sigma_r}$.}
    \label{fig: exciton layout}
\end{figure}

Fig.~\ref{fig: exciton layout} depicts the layout of these regions.
We will label the effective area of each region as $A$, the density of the localization regions as $\sigma_r$, and the actual excitonic density as $\sigma$. The fact that each region contains either 0 or 1 exciton implies that $\sigma \leq \sigma_r$ and that the occupation probability for any region is $\frac{\sigma}{\sigma_r}$.  For an exciton in the orbital state $\beta$ localized to a region centered at a generic position $\bm{R''}$, we will use a stationary wavepacket $f_{\bm{R''}}$ centered at $\bm{R''}$ to model the zero-point momentum excitonic center-of-mass, leading to the following composite wavefunction:
\begin{equation} \label{eq: exciton wavefunction 2}
\begin{gathered}
\Psi_{\beta,\bm{0}}(\bm{r_{CM}},\bm{d}) = f_{\bm{R''}}(\bm{r_{CM}}) \psi_{\beta}(d).
\end{gathered}
\end{equation}
We will label the center of the wavepacket for the unprimed (primed) exciton as $\bm{R}$ ($\bm{R'}$). The lack of overlap between neighboring wavepackets, as well as the separability of the center-of-mass part and the orbital part, imply that the matrix elements for $\bm{\Delta r}$, $\bm{d}$, and $\bm{d'}$ can be determined by expanding these variables in different Hilbert subspaces, i.e. the center-of-mass position basis for the two excitons for $\bm{\Delta r}$, the orbital states of the unprimed exciton for $\bm{d}$ , and the orbital states of the primed exciton for $\bm{d'}$.

\section{Interaction Hamiltonian Matrix Elements} \label{sec: Matrix Elements and Interaction Hamiltonian}
\par
In deriving the matrix elements, we will restrict the range of our summation to bright excitonic states centered at zero center-of-mass momentum. It is worth explaining how such a reduction is physically valid. Given two generic excitons, there are three possible interaction processes that can occur: annihilation (which preserves c.m. momentum), direct scattering (which alters c.m. momentum), or exchange. Though direct scattering has recently been theoretically analyzed \cite{ExcitonExcitonScattering}, it is forestalled if both of the excitons are initially bright and in the ground states of their respective orbital ladders. This is because the scattering process must lower the energy of one exciton while increasing the energy of the other, and since both excitons are at the minimum possible energy level, neither can decrease in energy while still maintaining coherence. 
\par
In addition, the large spacing between the excitons ensures that the decay rate is unaffected by exchange. Single-particle exchange (electron or hole) is hindered by the fact that the ground-state orbital restricts the electron-hole separation to a distance far closer than the separation between the localization region centers. Similarly, two-particle (full-exciton) exchange is hindered by the negligible overlap between the wavepackets corresponding even to neighboring localization regions.
\par
The only relevant exciton-exciton decay process is thus Auger recombination, for which center-of-mass momentum conservation requires that the non-annihilated exciton be excited to another bright state. Due to the large distance between neighboring wavepacket centers relative to the exciton size and the lack of overlap between the wavepackets, the $\bm{\Delta r}$ operator simplifies to the corresponding C-number, i.e. the displacement between the coordinate centers of the respective excitons. Henceforth, we will use $\bm{\Delta R}$ to label the constant $\bm{R} - \bm{R'}$.
\par
Next, we consider the expansion of the hole-to-electron vectors $\bm{d}$ and $\bm{d'}$ in the basis of the ladder of orbitals for their respective excitons. Knowing that each exciton is initially in the ground state $\ket{\nu}$ and summing over all possible final bound states $(\ket{\nu_f})$, unbound states $(\ket{\phi_{\bm{q}}})$, and the vacuum state $(\ket{fs})$, $\bm{d} \ket{\nu} \bra{\nu}$ takes the following form:
\begin{equation} \label{eq: hole-to-electron vector expansion}
\begin{split}
&\bm{d} \ket{\nu} \bra{\nu} \\
&= \bigg(\sum_{\nu_f} \ket{\nu_f}\bra{\nu_f} + \sum_{\bm{q}} \ket{\phi_{\bm{q}}}\bra{\phi_{\bm{q}}} + \ket{fs}\bra{fs} \bigg) \bm{d} \ket{\nu}\bra{\nu} \\
&= \braket{fs|\bm{d}|\nu} B_{\nu} + \sum_{\nu_f} \braket{\nu_f|\bm{d}|\nu} B^{\dag}_{\nu_f}B_{\nu} \\
&\quad + \sum_{\bm{q}} \braket{\phi_{\bm{q}}|\bm{d}|\nu}  D^{\dag}_{\bm{q}}B_{\nu}.
\end{split}
\end{equation}
Here, $B_{\nu''}$ denotes the annihilation operator corresponding to the bound state $\ket{\nu''}$, whereas $D_{\bm{q''}}$ represents the annihilation operator for the unbound state $\ket{\phi_{\bm{q''}}}$. We consider the first matrix element in this expression, which couples the ground excitonic state with the vacuum state, by decomposing the excitonic state into the electronic band states of the lattice, per the Keldysh formalism. Specifically, a bright excitonic state $\ket{\nu''}$ can be conceptualized as a superposition across wavevectors $\bm{k''}$ of composite electron-hole states, with the electron deriving from the lowest conduction band and the hole from the same wavevector at the highest valence band. The wavevectors are weighted by the Fourier envelope function $\psi_{\nu''}(\bm{k''})$, which has been previously derived for MoS\textsubscript{2} \cite{HaugKoch}:
\begin{equation}
\begin{gathered}
\ket{\nu''} = \sum_{\bm{k''}} \psi_{\nu''}(\bm{k''}) \ket{c(\bm{k''})v(\bm{k''})}.
\end{gathered}
\end{equation}
Using this representation, we calculate the matrix element representing the transition from the initial excitonic state $\ket{\nu}$ to the vacuum state $\ket{fs}$: 
\begin{equation}
\begin{gathered}
\braket{fs|\bm{d}|\nu} = \sum_{\bm{k}} \psi_{\nu}(\bm{k}) \braket{fs|\bm{d}|c(\bm{k})v(\bm{k})}.
\end{gathered}
\end{equation}
Physically, the process described by the right-hand-side inner product corresponds to the electron at $\ket{c(\bm{k})}$ dropping to $\ket{v(\bm{k})}$ and annihilating the hole there \cite{OpticalNonlinearitiesMoS2}:
\begin{equation}
\begin{gathered}
\braket{fs|\bm{d}|\nu} = \sum_{\bm{k}} \psi_{\nu}(\bm{k}) \braket{v(\bm{k})|\bm{r_e}|c(\bm{k})}.
\end{gathered}
\end{equation}
Next, we examine the matrix element connecting the ground state to another bound excitonic state $\ket{\nu_f}$. The most convenient and physically intuitive method for solving this is by expanding in the spatial basis $\bm{d}$ instead of in the band basis:
\begin{equation}
\begin{gathered}
\braket{\nu_f|\bm{d}|\nu} = \int_{lat} d^2d \psi^*_{\nu_f}(d) \bm{d} \psi_{\nu}(d).
\end{gathered}
\end{equation}
Finally, we analyze the matrix element connecting $\ket{\nu}$ to an unbound state denoted by $\ket{\phi_{\bm{q}}}$. This state can be considered as a composite of the free electron plane wave with momentum $\hbar\bm{q}$ and the free hole plane wave with momentum $-\hbar\bm{q}$ (the opposite values of the electron and hole momenta derive from the fact that the center-of-mass momentum must equal 0):
\begin{equation}
\begin{gathered}
\phi_{\bm{q}}(\bm{d}) = \frac{1}{\sqrt{A}} e^{i\bm{q} \cdot \bm{r_e}} e^{-i\bm{q} \cdot \bm{r_h}}
= \frac{1}{\sqrt{A}} e^{i\bm{q} \cdot \bm{d}}.
\end{gathered}
\end{equation}
Similarly, we calculate the matrix element representing the transition from the ground state to a generic unbound state by expanding in $\bm{d}$:
\begin{equation}
\begin{split}
\braket{\phi_{\bm{q}}|\bm{d}|\nu} &= \int_{lat} d^2d \phi^*_{\bm{q}}(\bm{d}) \bm{d} \psi_{\nu}(d) \\
&= \frac{1}{\sqrt{A}} \int_{lat} d^2d e^{-i\bm{q} \cdot \bm{d}} \bm{d} \psi_{\nu}(d).
\end{split}
\end{equation}
Since one of the excitons drops to the Fermi sea from the lowest excitonic state in the annihilation process, the other exciton must be excited by an energy equaling the gap between the lowest excitonic state and vacuum. Based on our tight-binding calculations, we know that the gap between conduction and valence band at the K/K'-points (which equals the maximum bound exciton energy) is approximately 2.2 eV, and it has been shown that the electron-hole binding energy reduces the ground state exciton energy to approximately 1.9 - 2.1 eV \cite{ExcitonEnergy1,ExcitonEnergy2}. Therefore, if one ground state exciton is annihilated, then the other must rise to an energy of 3.8 - 4.2 eV, thus exceeding the maximum bound exciton energy and creating an unbound exciton. Applying the exciton-exciton interaction energy derived in Eq.~\eqref{eq: exciton-exciton potential energy} to a pair of excitons in the lowest excitonic state and substituting the operator representations of $\bm{d}$ and $\bm{d'}$ derived above, we obtain the following column of matrix elements for the perturbation $H_{ex-ex}$ to the Hamiltonian due to exciton-exciton interaction:
\begin{widetext}
\begin{equation} \label{eq: exciton-exciton Hamiltonian}
\begin{split}
H_{ex-ex} \ket{\nu,\nu} \bra{\nu,\nu} &= 
-\frac{C}{(\Delta R)^3} \sum_{\bm{q}} \Bigg[\bigg(\frac{3 \bm{\Delta R} \cdot \braket{fs|\bm{d}|\nu}}{(\Delta R)^2} \bm{\Delta R} - \braket{fs|\bm{d}|\nu} \bigg) \cdot
\braket{\phi_{\bm{q}}|\bm{d}|\nu} B_{\nu} D'^{\dag}_{\bm{q}} B'_{\nu} \\
&\quad + \bigg(\frac{3 \bm{\Delta R} \cdot \braket{\phi_{\bm{q}}|\bm{d}|\nu}}{(\Delta R)^2} \bm{\Delta R} - \braket{\phi_{\bm{q}}|\bm{d}|\nu} \bigg) \cdot \braket{fs|\bm{d}|\nu} D^{\dag}_{\bm{q}} B_{\nu} B'_{\nu} \Bigg] \delta_{\omega_{\bm{q}},2\omega_{\nu}}.
\end{split}
\end{equation}
\end{widetext}
Here, $\hbar \omega_{\nu}$ denotes the energy of the lowest excitonic state $\ket{\nu}$, while $\hbar \omega_{\bm{q}}$ denotes the energy of the generic unbound state $\ket{\phi_{\bm{q}}}$. Note that this summation consists of two parts: one corresponding to the annihilation of the unprimed exciton and excitation of the primed exciton to an unbound state, and the other corresponding to the converse process. Henceforth, we will reduce the summation over $\bm{q}$ to only include the wavevectors which satisfy the condition $\omega_{\bm{q}} = 2\omega_{\nu}$ as required by the energy conservation condition encapsulated by the Kronecker delta function. Also note that we have replaced $\bm{d'}$ with $\bm{d}$ in the inner products, since the hole-to-electron position operator acts on the states of the corresponding exciton in the same manner.

\section{Calculating the Annihilation Rate} \label{sec: Calculating the Annihilation Rate}
\par
We calculate the overall annihilation rate for a single exciton as a function of spatial excitonic density using Fermi's Golden Rule \cite{FermiGoldenRule}. We assume that the excitonic localization regions form a closely-packed (triangular) lattice resembling the layout of the TMD itself, though it is worth noting that the overall annihilation rate for a grid (square) layout would differ by less than 15 percent. The spacing $l_0$ between the centers of nearest-neighbor regions relates to the area density of regions (i.e. the saturation excitonic density) $\sigma_r$ as follows:
\begin{equation}
\begin{gathered}
\sigma_r = \frac{2}{l_0^2\sqrt{3}}. 
\end{gathered}
\end{equation}
In order to determine the total annihilation rate for a given exciton, we will use series summation over the annihilation rates of the exciton upon interaction with all possible primed excitons (i.e. all other excitonic locations). We start by deriving the transition amplitude for a single exciton-exciton axis $\bm\hat{l}$, where $\bm{\Delta r} = l\bm{\hat{l}}$, using the Hamiltonian column from Eq.~\eqref{eq: exciton-exciton Hamiltonian}:
\begin{equation}
\begin{gathered}
\braket{fs,\phi_{\bm{q}}|H_{ex-ex}|\nu,\nu} = \\
-\frac{C}{l^3} \Big(3 \braket{fs|\bm{\hat{l}} \cdot \bm{d}|\nu} \braket{\phi_{\bm{q}}|\bm{\hat{l}} \cdot \bm{d}|\nu} - \braket{fs|\bm{d}|\nu} \cdot \braket{\phi_{\bm{q}}|\bm{d}|\nu} \Big).
\end{gathered}
\end{equation}
The rate $\Gamma_{single,drop}(l,\bm{q})$ at which the unprimed exciton drops from the lowest excitonic state $\ket{\nu}$ to the Fermi sea upon interacting with a single primed exciton that jumps to a state $\ket{\phi_{\bm{q}}}$ is calculated through Fermi's Golden Rule:
\begin{equation} \label{eq: gamma single drop single q}
\begin{gathered}
\Gamma_{single,drop}(l,\bm{q}) = \frac{2\pi}{\hbar} \rho({\bm{q}}) \Big|\braket{fs,\phi_{\bm{q}}|H_{ex-ex}|\nu,\nu}\Big|^2. \\
\end{gathered}
\end{equation}
Here, $\rho(\bm{q})$ denotes the unbound excitonic density of states at the orbital wavevector $\bm{q}$. Note that all unbound states $\ket{\phi_{\bm{q}}}$ for a given magnitude $q$ are degenerate. In order to find the total rate $\Gamma_{single,drop}(l)$ at which an exciton drops from $\ket{\nu}$ to the Fermi sea due to the interaction with another exciton at a distance $l$, we therefore need to sum over the transition rates to all possible final wavevectors $\bm{q}$ on a ring of radius $q$. This is equivalent to replacing the density of states at a single value $\bm{q}$ with the total density of states for the ring, which we label $\rho(E_q)$, and averaging the amplitude-squared term over all possible directions $\bm{\hat{q}}$:
\begin{widetext}
\begin{equation} \label{eq: gamma single drop}
\begin{split}
\Gamma_{single,drop}(l) &= \frac{2\pi}{\hbar} \rho(E_q) \bigg<\Big|\braket{fs,\phi_{\bm{q}}|H_{ex-ex}|\nu,\nu}\Big|^2\bigg>_{\bm{\hat{q}}} \\
&= \frac{2\pi}{\hbar} \rho(E_q) \frac{C^2}{l^6} \bigg<\Big|3 \braket{fs|\bm{\hat{l}} \cdot \bm{d}|\nu} \braket{\phi_{\bm{q}}|\bm{\hat{l}} \cdot \bm{d}|\nu} + \braket{fs|\bm{d}|\nu} \cdot \braket{\phi_{\bm{q}}|\bm{d}|\nu} \Big|^2\bigg>_{\bm{\hat{q}}}.
\end{split}
\end{equation}
\end{widetext}
In addition, it is important to consider how the newly generated unbound exciton relaxes. Since the constituent electron and hole feature the same wavevector in reciprocal space, the electron can radiatively decay into the hole. However, due to the weak oscillator strength corresponding to the transition between conduction and valence bands at a particular wavevector (relative to the exciton-vacuum oscillator strength), this process is inefficient compared to the radiative decay of an exciton from the lowest excitonic state. A faster decay channel is by multiphonon relaxation, as previously analyzed in linear molecular aggregates \cite{LinearAggregates1, LinearAggregates2}.
\begin{figure}
    \centering
    \includegraphics[width=\columnwidth]{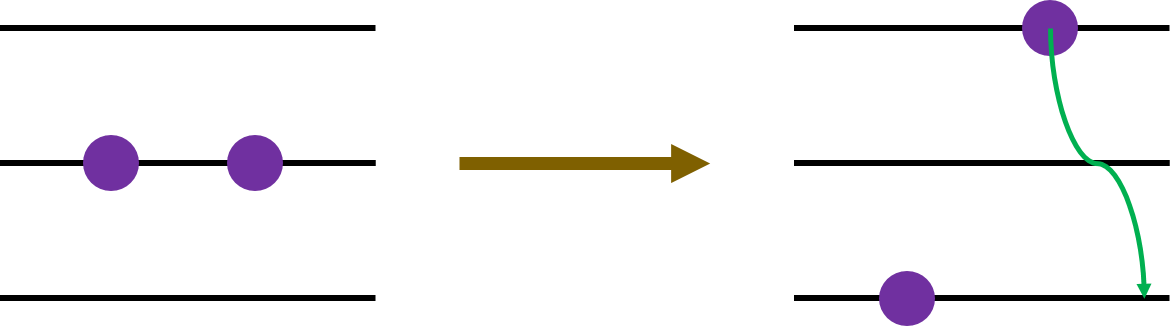}
    \caption{Simplified 3-level diagram representing the exciton-exciton annihilation process. Note that the excited exciton decays to the vacuum state upon interacting with phonons, as shown by the curved green arrow.}
    \label{fig: exciton three level}
\end{figure}
Due to the very small value of the transition dipole moment connecting the lowest excitonic state to a generic unbound state \cite{OpticalNonlinearitiesMoS2}, the unbound exciton is eventually annihilated to vacuum by such a process (as depicted in Fig.~\ref{fig: exciton three level}) instead of returning to the original lowest excitonic state. As such, the actual annihilation rate for a single exciton upon interaction with another exciton, $\Gamma_{single}(l)$, is double that of $\Gamma_{single,drop}(l)$:
\begin{widetext}
\begin{equation} \label{eq: gamma single}
\begin{gathered}
\Gamma_{single}(l) = \frac{4\pi}{\hbar} \rho(E_q) \frac{C^2}{l^6} \bigg<\Big|3 \braket{fs|\bm{\hat{l}} \cdot \bm{d}|\nu} \braket{\phi_{\bm{q}}|\bm{\hat{l}} \cdot \bm{d}|\nu} - \braket{fs|\bm{d}|\nu} \cdot \braket{\phi_{\bm{q}}|\bm{d}|\nu} \Big|^2\bigg>_{\bm{\hat{q}}}.
\end{gathered}
\end{equation}
\end{widetext}
The density of states $\rho(E_q)$ can be derived by envisioning a circular ring of free-particle states in a 2-dimensional reciprocal space. In this space, each state occupies a reciprocal area of $\frac{(2\pi)^2}{A}$. The fact that each of the two charge carriers (electron and hole) is propagating at a wavevector of magnitude $q$ results in the following relationship between the unbound state energy and $q$:
\begin{equation}
\begin{gathered}
E_q = E_{gap} + \frac{\hbar^2 q^2}{2} \bigg(\frac{1}{m_e} + \frac{1}{m_h}\bigg).
\end{gathered}
\end{equation}
Here, $E_{gap}$ represents the lattice band gap energy, equaling the sum of the ground state excitonic energy and the exciton binding energy. Defining the region of the reciprocal space enclosed by a circle of radius $q$ as $a_q = \pi q^2$, we calculate the density of states as follows:
\begin{equation} \label{eq: density of states}
\begin{split}
\rho(E_q) &= \frac{dN}{da_q} \frac{da_q}{dq} \frac{dq}{dE_q} \\
&= \frac{A}{(2\pi)^2} 2 \pi q \frac{1}{\hbar^2 q} \bigg(\frac{1}{m_e} + \frac{1}{m_h}\bigg)^{-1} \\
&= \frac{m_e m_h}{2 \pi \hbar^2 (m_e + m_h)} A.
\end{split}
\end{equation}
Next, we determine the total annihilation rate for a given exciton. Note that the probability of interaction between two excitons separated by a distance $l$ is proportional to $l^{-6}$, and therefore for a given exciton, the interaction with the nearest-neighbor excitons should dominate. Establishing the unprimed exciton as the origin of a 2D coordinate system, we label the reduced coordinates for a generic localization region as $(n,m)$, where $n$ and $m$ are integers and the horizontal and vertical coordinates ($h$ and $v$, respectively) equal the following:
\begin{equation}
\begin{split}
h_{n,m} &= n + m\cos{\frac{\pi}{3}} = n + \frac{1}{2}m, \\
v_{n,m} &= m\sin{\frac{\pi}{3}} = \frac{\sqrt{3}}{2}m.
\end{split}
\end{equation}
\textit{For the case in which every localization region is occupied by an exciton}, the distance between nearest-neighbor excitons equals $l_0$. We aim to obtain the ratio between the total annihilation rate for the unprimed exciton to the annihilation rate with a single nearest-neighbor primed exciton. To do so, we use the following summation over the localization regions indexed by $(n,m)$:
\begin{equation} \label{eq: gamma single to gamma total summation}
\begin{split}
\sum_{n,m} \bigg(\frac{l_0}{l_{n,m}}\bigg)^6 &= \sum_{n,m} \frac{1}{\bigg(\Big(n + \frac{1}{2}m\Big)^2 + \Big(\frac{\sqrt{3}}{2}m\Big)^2\bigg)^3} \\
&= \sum_{n,m} \frac{1}{(n^2 + nm + m^2)^3}.
\end{split}
\end{equation}
We simplify this sum by only applying it to a single $\frac{\pi}{3}$ slice. Due to the 6-fold rotational symmetry of the excitonic layout, the result for a single slice will apply to every other quadrant and axis, respectively. For the first slice (starting from the positive horizontal axis, inclusive, and ending just short of the $\frac{\pi}{3}$ axis), we obtain $\sum_{n=1}^{\infty} \sum_{m=0}^{\infty} \frac{1}{(n^2 + nm + m^2)^3} \approx 1.0626$. The result of Eq.~\eqref{eq: gamma single to gamma total summation} is calculated by multiplying these results by 6:
\begin{equation} \label{eq: gamma single to gamma total ratio}
\begin{gathered}
\sum_{n,m} \frac{1}{(n^2 + nm + m^2)^3} \approx 6 \times 1.0626
\approx 6.4.
\end{gathered}
\end{equation}
We determine the total decay rate for a single exciton \textit{given full filling of the localization regions}, $\Gamma_{total,full}$, by multiplying the expression from Eq.~\eqref{eq: gamma single} by the ratio from Eq.~\eqref{eq: gamma single to gamma total ratio} and substituting the density of states from Eq.~\eqref{eq: density of states}, while averaging over all possible exciton-exciton axis orientations $\bm\hat{l}$:
\\
\begin{widetext}
\begin{equation} \label{eq: gamma total full}
\begin{split}
\Gamma_{total,full}
&= \frac{12.8 m_e m_h}{\hbar^3 (m_e + m_h)} \frac{C^2 A}{l_0^6} 
\bigg<\Big|3 \braket{fs|\bm{\hat{l}} \cdot \bm{d}|\nu} \braket{\phi_{\bm{q}}|\bm{\hat{l}} \cdot \bm{d}|\nu} - \braket{fs|\bm{d}|\nu} \cdot \braket{\phi_{\bm{q}}|\bm{d}|\nu} \Big|^2\bigg>_{\bm{\hat{l}},\bm{\hat{q}}} \\
&= \frac{8.3 m_e m_h}{\hbar^3 (m_e + m_h)} C^2 A \sigma_r^3 
\bigg<\Big|3 \braket{fs|\bm{\hat{l}} \cdot \bm{d}|\nu} \braket{\phi_{\bm{q}}|\bm{\hat{l}} \cdot \bm{d}|\nu} - \braket{fs|\bm{d}|\nu} \cdot \braket{\phi_{\bm{q}}|\bm{d}|\nu} \Big|^2\bigg>_{\bm{\hat{l}},\bm{\hat{q}}}.
\end{split}
\end{equation}
\end{widetext}
For the general case including full or partial filling, i.e. $\sigma \leq \sigma_r$, Eq.~\eqref{eq: gamma total full} must be weighted by the occupation probability of any given localization region, i.e. $\frac{\sigma}{\sigma_r}$. This yields an annihilation rate $\Gamma_{total}$ for a single exciton that varies linearly with the excitonic density $\sigma$:
\begin{widetext}
\begin{equation} \label{eq: gamma total}
\begin{split}
\Gamma_{total}(\sigma) &= \frac{\sigma}{\sigma_r} \Gamma_{total,full} \\
&= \frac{8.3 m_e m_h}{\hbar^3 (m_e + m_h)} C^2 A \sigma_r^2 \sigma
\bigg<\Big|3 \braket{fs|\bm{\hat{l}} \cdot \bm{d}|\nu} \braket{\phi_{\bm{q}}|\bm{\hat{l}} \cdot \bm{d}|\nu} - \braket{fs|\bm{d}|\nu} \cdot \braket{\phi_{\bm{q}}|\bm{d}|\nu} \Big|^2\bigg>_{\bm{\hat{l}},\bm{\hat{q}}}.
\end{split}
\end{equation}
\end{widetext}

\section{Results and Discussion} \label{sec: Results and Discussion}
\par
\begin{figure}
    \centering
    \includegraphics[width=\columnwidth]{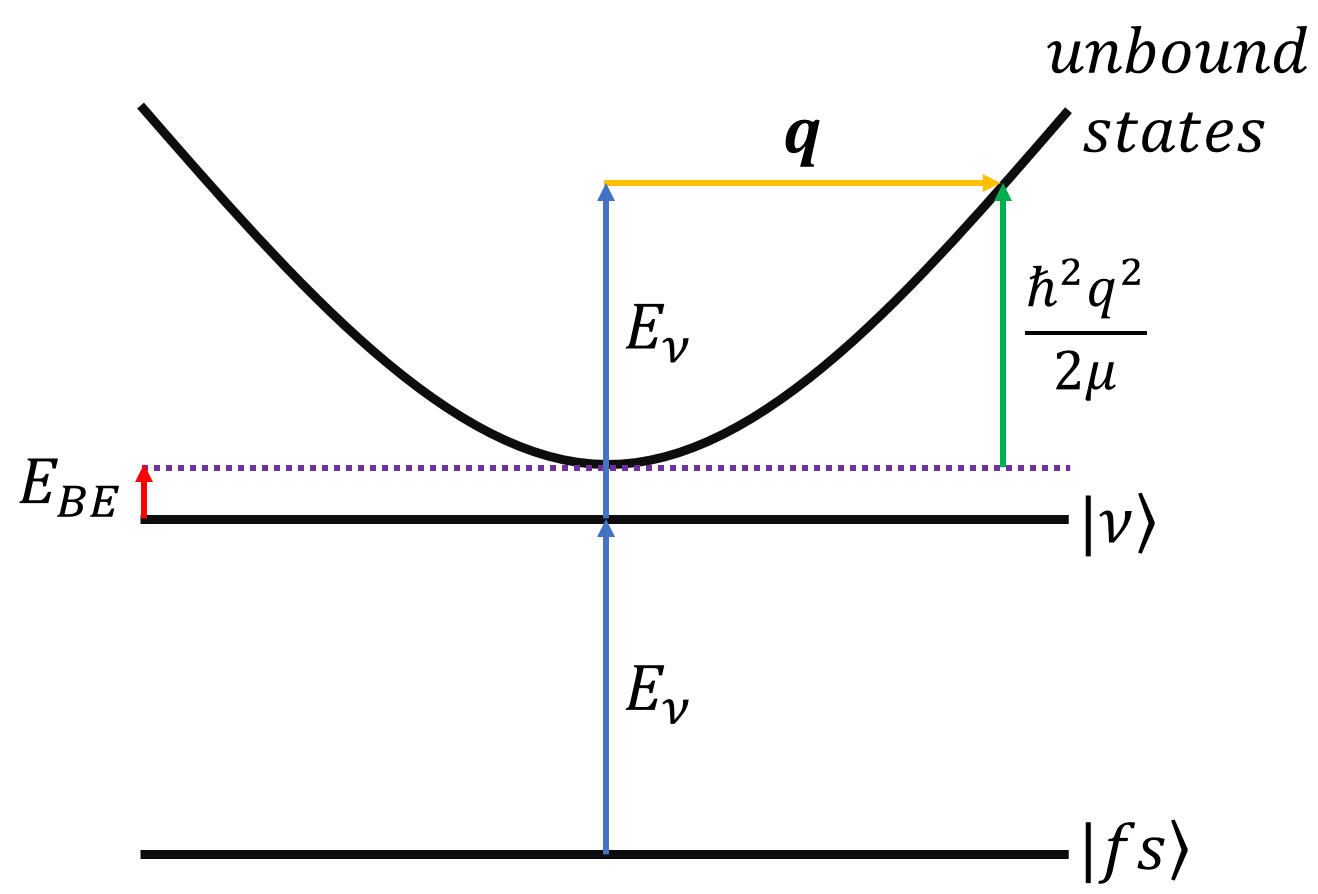}
    \caption{Depiction of excitonic energy levels. The energy gap between the excitonic ground state $\ket{\nu}$ and the excited unbound state at wavevector $\bm{q}$ equals the gap $E_{\nu}$ between $\ket{\nu}$ and the vacuum state $\ket{fs}$. Note that $\mu$ represents the reduced mass of the electron-hole pair, given by $\Big(\frac{1}{m_e} + \frac{1}{m_h}\Big)^{-1} \approx \frac{m}{2}$, where $m \approx m_e \approx m_h$.}
    \label{fig: exciton levels}
\end{figure}

The final step in the remaining process of obtaining the exciton-exciton annihilation rate is to substitute numerical values into the matrix elements and coefficients in Eq.~\eqref{eq: gamma total}. The matrix element $\braket{fs|\bm{d}|\nu}$, corresponding to the annihilation of an exciton at the ground state $\ket{\nu}$, was previously determined (via decomposition into the band basis \cite{OpticalNonlinearitiesMoS2}) as proportional to the square root of the localization region area $A$:
\begin{equation} \label{eq: ground state to Fermi sea}
\begin{gathered}
\braket{fs|\bm{d}|\nu} \approx \bm{\hat{x}} (0.1425i) \sqrt{A}.
\end{gathered}
\end{equation}
Here, we have defined $\bm{\hat{x}}$ as the direction of the vector $\braket{fs|\bm{d}|\nu}$. Another element used in Eq.~\eqref{eq: gamma total} is $\braket{fs|\bm{\hat{l}} \cdot \bm{d}|\nu}$, i.e. the $\hat{l}$-component of $\braket{fs|\bm{d}|\nu}$. Labeling the angle between $\bm{\hat{x}}$ and $\bm{\hat{l}}$ as $\phi_l$, we obtain the following expression for that inner product:
\begin{equation}
\begin{split}
\braket{fs|\bm{\hat{l}} \cdot \bm{d}|\nu} &= \bm{\hat{l}} \cdot \braket{fs|\bm{d}|\nu} \\
&\approx (0.1425i) \sqrt{A} \cos{\phi_l}.
\end{split}
\end{equation}
We calculate the inner product $\braket{\phi_{\bm{q}}|\bm{d}|\nu}$, corresponding to the excitation of a ground state exciton to the unbound state exhibiting an energy twice that of the ground state, by expanding in the hole-to-electron vector position basis and integrating:
\begin{align}
\begin{split}
\braket{\phi_{\bm{q}}|\bm{d}|\nu} &= \frac{1}{\sqrt{A}} \int_{lat} d^2d e^{-i\bm{q} \cdot \bm{d}} \bm{d} \psi_{\nu}(d) \\
&= -\bm{\hat{q}} \frac{8i \sqrt{2\pi} a_0^3 q}{(a_0^2 q^2 + 4)^2 \sqrt{A}}.
\end{split}
\end{align}
Here, $\frac{a_0}{2}$ represents the expectation value of the electron-hole distance, which approximately equals 0.67 nm for MoS\textsubscript{2} \cite{OpticalNonlinearitiesMoS2}. Labeling the angle between $\bm{\hat{x}}$ and $\bm{\hat{q}}$ as $\phi_q$, the $\hat{l}$-component of the inner product is solved as follows:
\begin{equation}
\begin{split}
\braket{\phi_{\bm{q}}|\bm{\hat{l}} \cdot \bm{d}|\nu} &= \bm{\hat{l}} \cdot \braket{\phi_{\bm{q}}|\bm{d}|\nu} \\
&= -i \frac{8 \sqrt{2\pi} a_0^3 q}{(a_0^2 q^2 + 4)^2 \sqrt{A}} \cos{(\phi_l - \phi_q)}.
\end{split}
\end{equation}
Since $\bm{\hat{l}}$ and $\bm{\hat{q}}$ can take any direction relative to $\bm{\hat{x}}$, we need to average over $\phi_l$ and $\phi_q$. Substituting the matrix elements into Eq.~\eqref{eq: gamma total}, and averaging over the two angle variables, we obtain the following expression for the total annihilation rate as a function of the excitonic density $\sigma$:
\begin{widetext}
\begin{equation} \label{eq: gamma total 2}
\begin{split}
\Gamma_{total}(\sigma) 
&= \frac{8.3 m_e m_h}{\hbar^3 (m_e + m_h)} C^2 A \sigma_r^2 \sigma \Bigg|\frac{(-8i) (0.1425i) \sqrt{2\pi} a_0^3 q}{(a_0^2 q^2 + 4)^2} \Bigg|^2 
\bigg<\bigg<\Big|3\cos{\phi_l}\cos{(\phi_l - \phi_q)} - \cos{\phi_q}\Big|^2\bigg>_{\phi_l}\bigg>_{\phi_q} \\
&= \frac{27 \pi a_0^6 q^2}{(a_0^2 q^2 + 4)^4} \frac{\mu}{\hbar^3} C^2 A \sigma_r^2 \sigma.
\end{split}
\end{equation}
\end{widetext}
Here, $\mu$ denotes the reduced mass of the electron-hole pair, defined as $\mu = (1/m_e + 1/m_h)^{-1}$. Fig.~\ref{fig: exciton levels} provides a schematic of the relevant excitonic energy levels (Fermi sea, ground state, and unbound states) and the gaps between those levels. Recall that the excited exciton gains an energy of $E_{\nu}$, equaling that of the ground state exciton relative to the Fermi sea. Some of this energy is spent in overcoming the exciton binding energy $E_{BE}$ in order to reach the conduction band energy, while the remainder is used in reaching the unbound state $\ket{\phi_{\bm{q}}}$, which exhibits an energy (relative to the conduction band) corresponding to a free electron and free hole carrying wavevectors of magnitude $q$.
In quantitative terms, the binding energy, wavevector magnitude, and ground state energy are related by the following expression:
\begin{equation} \label{eq: state energy expression}
\begin{gathered}
E_{BE} + \frac{\hbar^2 q^2}{2\mu} = E_{\nu}.
\end{gathered}
\end{equation}
Note that the annihilation rate varies with the reduced mass $\mu$ (which is defined by the effective electron and hole masses and is always smaller than the smallest effective mass of composite particles), as evidenced from Eqs.~\eqref{eq: gamma total 2} and~\eqref{eq: state energy expression}. In a previous study \cite{OpticalNonlinearitiesMoS2}, we determined $m_e$ and $m_h$ each as 0.55 times the actual electron mass $m_0$, $E_{BE}$ as 0.31 eV, and $E_{\nu}$ as 1.92 eV. It is worth mentioning that values for $m_h$ and $m_e$ equaling $0.65m_0$ and $0.49m_0$, respectively, have been determined in other analysis \cite{BandCurvature}, yielding the same reduced mass as the case in which both carriers have a mass of $0.55m_0$. Substituting these values into the above expression, we find that $q = 3.4 \textrm{ nm}^{-1}$. Given this value for $q$, along with $a_0 = 1.34\textrm{ nm}$ and $m_e \approx m_h \approx 5.0 \times 10^{-31}\textrm{ kg}$, our last computational step is to obtain $C$. This constant simply equals $\frac{e^2}{4 \pi \epsilon_0 \epsilon_r}$, where $\epsilon_r$ represents the ambient dielectric constant for the monolayer MoS\textsubscript{2} film. For a free-standing film, the dielectric constant equals 1, yielding $C = 2.3 \times 10^{-28} \textrm{ } \frac{\textrm{kg} \textrm{m}^3}{\textrm{s}^2}$ and reducing the annihilation rate from Eq.~\eqref{eq: gamma total 2} to a function of effective sample area, density of regions, and exciton density:
\begin{equation} \label{eq: gamma total 4}
\begin{gathered}
\Gamma_{ex-ex}(\sigma) \approx \bigg(1.7 \times 10^{-22} \textrm{ } \frac{\textrm{m}^4}{\textrm{s}}\bigg) A \sigma_r^2 \sigma.
\end{gathered}
\end{equation}
Note that this represents the decay rate for a single exciton. From this expression, we derive the rate of change of density due to the annihilation process, labeling the total sample area being measured as $A_{total}$ and the total number of excitons as $N$:
\begin{equation}
\begin{split}
\frac{d \sigma}{dt} = \frac{1}{A_{total}} \frac{dN}{dt} &= -\frac{N}{A_{total}} \Gamma_{ex-ex} \\
&= -\alpha \sigma^2.
\end{split}
\end{equation}
Here, $\alpha$ represents the annihilation rate constant, given by the following value:
\begin{equation} \label{eq: rate constant raw}
\begin{gathered}
\alpha \approx \bigg(1.7 \times 10^{-22} \textrm{ } \frac{\textrm{m}^4}{\textrm{s}}\bigg) A \sigma_r^2.
\end{gathered}
\end{equation}
It is worth determining a feasible range for $\alpha$ by analyzing $A$ and $\sigma_r$. As previously mentioned, $A$ represents the localization area of each charge carrier wavefunction, while $\sigma_r$ represents the exciton saturation density. Assuming that the localization regions approximately fill the position space of the sample, we find that $\sigma_r \approx \frac{1}{A}$, yielding the following value for $\alpha$ in terms of $A$:
\begin{equation} \label{eq: rate constant}
\begin{gathered}
\alpha \approx \bigg(1.7 \times 10^{-22} \textrm{ } \frac{\textrm{m}^4}{\textrm{s}}\bigg) \frac{1}{A}.
\end{gathered}
\end{equation}
Next, we aim to find a feasible range for $A$, starting with a consideration of the defects that give rise to exciton and charge carrier localization. Since the grain boundaries of graphene were mapped by Kim \textit{et al.} \cite{GrapheneGrainBoundaries}, the characterization of inhomogeneities in 2D materials has remained an active field. To this end, the high concentration of defects in TMD samples obtained by mechanical exfoliation or grown epitaxially has been well established by experimental measurements. For mechanically exfoliated monolayer MoS\textsubscript{2}, Wang \textit{et al.} \cite{MoS2DefectStructure1} determined a defect density ranging from $0.3 \times 10^{12}$ to $2 \times 10^{13} \textrm{ cm}^{-2}$ through a pump-probe measurement, while Vancso \textit{et al.} \cite{MoS2DefectStructure2} used a similar method to obtain a defect density of $5 \times 10^{12}$ to $5 \times 10^{13} \textrm{ cm}^{-2}$. Similarly, Liu \textit{et al.} \cite{MoS2DefectStructure3} measured an average defect density of $8 \times 10^{12} \textrm{ cm}^{-2}$ on monolayer MoS\textsubscript{2} grown on epitaxial graphene. CVD-grown TMDs exhibit an even greater density of inhomogeneities, although a recent study by Rogers \textit{et al.} \cite{LaserAnnealingCVD} has shown that laser annealing can remove some of the impurities and reduce the strain gradient.
\par
The range of defect densities gives rise to an array of possible excitonic coherence lengths. For a particular region of a sample, the local value of $A$ can be inferred via the fundamental $A$-dependence of the excitonic radiative decay rate. We use the following well-known expression based on the Einstein coefficients \cite{EinsteinCoefficients} to calculate the radiative loss rate from the excitonic state $\ket{\nu}$ to the vacuum state $\ket{fs}$:
\begin{equation}
\begin{gathered}
\Gamma_{rad} = \frac{\omega^3}{3 \pi \epsilon_0 \hbar c^3} \Big|\braket{fs|e\bm{d}|\nu}\Big|^2.
\end{gathered}
\end{equation}
Substituting the excitonic energy $\hbar \omega = 1.9 \textrm{ eV}$ and the dipole matrix element $|\braket{fs|\bm{d}|\nu}| = 0.1425 \sqrt{A}$ (see Eq.~\eqref{eq: ground state to Fermi sea}) into this expression, we find that the radiative decay rate varies linearly with the localization area $A$:
\begin{equation} \label{eq: radiative decay rate}
\begin{gathered}
\Gamma_{rad} = \bigg(5.0 \times 10^{25} \textrm{ } \frac{1}{\textrm{m}^2\textrm{s}}\bigg) A.
\end{gathered}
\end{equation}
Note that the linear relationship between the radiative decay rate and localization area agrees with Wang \textit{et al.} \cite{RadiativeDecayMoS2}. For excitons with large localization areas, the radiative lifetime drops to the femtosecond range and will be far shorter than the exciton-exciton annihilation lifetime (even observations of rapid annihilation in molybdenum-based TMDs yield lifetimes in the tens of picoseconds \cite{RapidExcitonAnnihilationMoS2, RapidExcitonAnnihilationMoSe2}), thus rendering the latter process negligibly slow. 
\par
Next, we resolve the range of localization areas that fit our assumption that each localization region contains up to one exciton. To this end, it is important to note that although delocalized excitons act in a bosonic manner, the behavior becomes fermion-like given significant localization (on the scale of the excitonic size). For example, Ohtsu \cite{FermionicExcitons} shows that for an sufficiently small exciton localization region indexed by $n$, the commutatator between the excitonic annihilation and creation operators takes the following form:
\begin{equation}
\begin{gathered}
[B_n,B^{\dag}_n] = 1 - 2 B^{\dag}_n B_n.
\end{gathered}
\end{equation}
The fact that the formation of more than one exciton in a single narrow localization region is not favored is further supported by experimental evidence of phase space filling, e.g. the absorption saturation measured in GaAs by Hunsche \textit{et al.} \cite{PhaseSpaceFilling}. For 2D materials, the seminal theoretical analysis by Schmitt-Rink \textit{et al.} \cite{ExcitonBlockedZone} showed that the effects of Pauli exclusion result in a blocked area of $8.5 \pi a^2$ around each exciton (where $a$ represents the exciton size) \cite{ExcitonBlockedZone2}, indicating a diameter of $5.8 a$ for the blocked zone. For molybdenum-based TMDs, the exciton size approximately equals $0.7 \textrm{ nm}$ as previously mentioned, leading to a blocked zone diameter of about $4 \textrm{ nm}$. This is confirmed by recent measurements on monolayer MoSe\textsubscript{2} by Kumar \textit{et al.} \cite{RapidExcitonAnnihilationMoSe2}, which suggest that the excitonic density saturates at a nearest-neighbor spacing of $4 \textrm{ nm}$. 
\par
\begin{figure}
    \centering
    \includegraphics[width=\columnwidth]{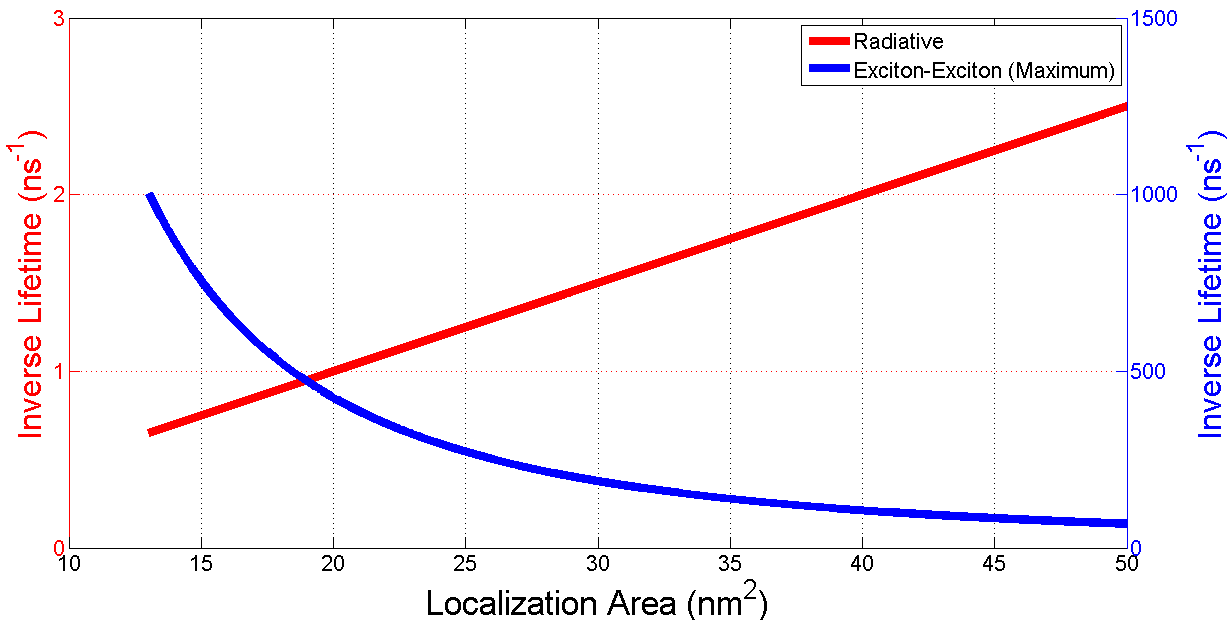}
    \caption{Plot of the decay rates for a single exciton due to radiative recombination (red) and exciton-exciton annihilation (blue) given maximal filling of the excitonic localization regions, as functions of the localization area $A$ for the range $13 \textrm{ nm}^2 \lesssim A \lesssim 50 \textrm{ nm}^2$. Note that the radiative recombination rate is more than an order of magnitude less than the exciton-exciton annihilation rate for all values of $A$ in this range.}
    \label{fig: decay rates}
\end{figure}

Given a diameter of less than about $8 \textrm{ nm}$ for each localization region, the formation of more than one exciton in a single region is hindered since a region of such size cannot fit more than one blocked zone. On the other hand, if the localization region diameter is less than the blocked zone diameter of about $4 \textrm{ nm}$, then the orbital wavefunction for an exciton formed in that region would start to experience distortion due to the effects of strong localization. As a result, our model of quasi-localized excitons behaving in a fermion-like manner within their respective regions specifically applies for the range of areas given by $13 \textrm{ nm}^2 \lesssim A \lesssim 50 \textrm{ nm}^2$. For this range, the density of regions $\sigma_r$ ($\approx 1/A$) represents the excitonic saturation density and thus serves as the upper bound for $\sigma$. As a result, the maximum value of the annihilation rate from Eq.~\eqref{eq: gamma total 4} (which applies when the localization regions are saturated with excitons) becomes a function solely of the localization area:
\begin{equation} \label{eq: maximum exciton-exciton annihilation rate}
\begin{gathered}
\Gamma_{ex-ex,max} \approx \bigg(1.7 \times 10^{-22} \textrm{ } \frac{\textrm{m}^4}{\textrm{s}}\bigg) \frac{1}{A^2}.
\end{gathered}
\end{equation}
Fig.~\ref{fig: decay rates} depicts the radiative recombination and maximum exciton-exciton annihilation rates based on Eqs.~\eqref{eq: radiative decay rate} and~\eqref{eq: maximum exciton-exciton annihilation rate}, respectively. 
For the entire range, the annihilation process given maximal excitonic density is more than an order of magnitude faster than the radiative process, with the effect especially pronounced at smaller localization areas. This implies that if a sufficient pump fluence is applied to the sample such that the localization regions become saturated, the initial radiative recombination rate will be negligible compared to the exciton-exciton annihilation rate.
\par
We now return to the general case in which the excitonic density $\sigma$ is decoupled from the density of localization regions $\sigma_r$ (i.e. the excitonic density is a variable with an arbitrary value less than or equal to the region density) in order to determine the range for the rate constant $\alpha$ for the relevant range of localization areas. From Eq.~\eqref{eq: rate constant}, it is apparent that $\alpha$ decreases with $A$. This value is maximized (minimized) when the localization area $A$ is minimized (maximized), which applies when $A \approx 13 \textrm{ nm}^2$ ($50 \textrm{ nm}^2$):
\begin{equation} \label{eq: rate constant range}
\begin{gathered}
3.4 \times 10^{-6} \textrm{ } \frac{\textrm{m}^2}{\textrm{s}} \lesssim \alpha \lesssim 1.3 \times 10^{-5} \textrm{ } \frac{\textrm{m}^2}{\textrm{s}}.
\end{gathered}
\end{equation}
It is useful to compare this result to experimental findings demonstrating rapid exciton-exciton annihilation. In this respect, the predicted range for the annihilation rate constant for quasi-localized excitons at low temperatures resembles values measured at room temperature. For example, Sun, Rao, \textit{et al.} \cite{RapidExcitonAnnihilationMoS2} determined a rate constant of $4.3 \times 10^{-6} \textrm{ } \textrm{m}^2/\textrm{s}$ for MoS\textsubscript{2}, well within the span shown in Eq.~\eqref{eq: rate constant range}. Similarly, Kumar \textit{et al.} \cite{RapidExcitonAnnihilationMoSe2} measured a rate constant of $3.3 \times 10^{-5} \textrm{ } \textrm{m}^2/\textrm{s}$, slightly exceeding our predicted range. Although it is difficult to glean the localization areas from the radiative decay rate in these experiments since the radiative lifetime measurements were performed on a thermal ensemble consisting of both bright and dark excitons, the similarities between our predictions and the experimental results suggest that the rate constant might be similar for localized and delocalized excitons. 
A similarity between the annihilation rates of localized and delocalized excitons would imply that the exchange interaction between delocalized excitons is much weaker than the dipole-dipole interaction that gives rise to annihilation in both the localized and delocalized cases.
\par
One possible means of testing our predictions is through an experimental method that exploits the narrow (micron-ranged) spot size of the pumping beam relative to the total area of the 2D sample. Due to the uneven distribution of imhomogeneities across the sample (as evidenced by the wide range of defect densities discussed above), the excitonic localization area should differ from one beam-sized region to the next. The localization area for each region can be gleaned by measuring the radiative decay rate for the region under cryogenic conditions. Of course, another method would be to test on samples synthesized using different techniques. As shown in Fig.~\ref{fig: decay rates}, our calculations will apply specifically when the radiative decay rate falls in the range $6.5 \times 10^8 \textrm{ s}^{-1} \lesssim \Gamma_{rad} \lesssim 2.5 \times 10^9 \textrm{ s}^{-1}$. It is worth noting that such long radiative lifetimes have already been observed in superacid-treated TMD samples \cite{Superacid1, Superacid2}. The best strategy for obtaining a consistent localization area across the sample, however, is likely to break up the lattice into quantum dots. Fabrication techniques for patterning quantum dots into monolayer TMDs have been rapidly advancing in terms of both precision and miniaturization, with Wei \textit{et al.} \cite{QuantumDots1} generating nanodots as small as $15 \textrm{ nm}$, and more recently Ding \textit{et al.} \cite{QuantumDots2} reducing the size to the single nanometer range. 

\section{Conclusion} \label{sec: Conclusion}
\par
We have derived the annihilation rate for optically generated excitons in monolayer MoS\textsubscript{2} as a function of density, given large inter-exciton spacing (significantly exceeding the excitonic size) and quasi-localization of individual excitons. To start, we derived the exciton-exciton interaction energy as a function of exciton spacing, using the dipole-dipole approximation to show that the energy is linear in the hole-to-electron vector for each exciton. We then promoted the hole-to-electron vectors to orbital ladder operators and demonstrated that the center-to-center vector approximately takes a single value, thus enabling the center-of-mass states to be abstracted out and reducing the dimensionality of the Hilbert space. Finally, we employed Fermi's Golden Rule to calculate the exciton-exciton annihilation rate, using energy conservation to determine the final state in the orbital ladder reached by the excited exciton. We obtained a rate that varies inversely with the localization area and linearly with the total excitonic density. The rates measured in experimental observations of rapid annihilation resemble our predicted range, even though the experimental data was obtained at room temperature. 
\par
Our findings present both theoretical and experimental applications. On the theoretical side, they can be expanded to other 2D materials by re-evaluating the numerical results for the matrix elements as well as the energy levels. On the experimental side, when conducting measurements of nonradiative decay rates on samples with low excitonic densities, the results of this paper can be used to account for the loss rate due to exciton-exciton annihilation. Most importantly, since the annihilation rate varies with exciton density, the process introduces a nonlinearity in the transient behavior of the excitons and knowledge of the rate thus represents a critical step toward assessing the utility of this material in the context of cavity nonlinear optics.

\begin{acknowledgements}

This work was supported by the National Science Foundation (NSF) through Awards PHY-1648807 and DMR-1838497, as well as by a seed grant from the Precourt Institute for Energy at Stanford University. Sandia National Laboratories is a multimission laboratory managed and operated by National Technology and Engineering Solutions of Sandia, LLC., a wholly owned subsidiary of Honeywell International, Inc., for the U.S. Department of Energy\textsc{\char13}s National Nuclear Security Administration under contract DE-NA-0003525.

\end{acknowledgements}

\end{document}